\documentclass[11pt,a4paper]{article} 
\pdfoutput=1

\usepackage{jcapmod} 
\usepackage{booktabs}
\usepackage{url}
\usepackage[utf8]{inputenc}

\newcommand{\eq}[1]{eq. (\ref{#1})} 
\newcommand{\sektion}[1]{section \ref{#1}}

\def\be{\begin{equation}}
\def\ee{\end{equation}}
\def\bea{\begin{eqnarray}}
\def\eea{\end{eqnarray}}

\title{Primordial monopoles, proton decay, gravity waves and GUT inflation}

\author[a]{Vedat Nefer \c{S}eno\u{g}uz}
\author[b]{and Qaisar Shafi}

\affiliation[a]{Department of Physics, Mimar Sinan Fine Arts University, 34380
\c{S}i\c{s}li, \.Istanbul, Turkey}
 \affiliation[b]{Bartol Research Institute, Department of Physics and Astronomy, University of Delaware, Newark, DE 19716, USA}

\abstract{We consider non-supersymmetric GUT inflation models in which
intermediate mass monopoles may survive inflation because of the restricted
number of e-foldings experienced by the accompanying symmetry breaking.  Thus,
an observable flux of primordial magnetic monopoles, comparable to or a few
orders below the Parker limit, may be present in the galaxy. The mass scale
associated with the intermediate symmetry breaking is $10^{13}$ GeV for an
observable flux level, with the corresponding monopoles an order of magnitude or
so heavier.  Examples based on $SO(10)$ and $E_6$ yield such intermediate mass
monopoles carrying respectively two and three units of Dirac magnetic charge.
For GUT inflation driven by a gauge singlet scalar field with a Coleman-Weinberg
or Higgs potential, compatibility with the Planck measurement of the scalar
spectral index yields a Hubble constant (during horizon exit of cosmological
scales) $H \sim 7$--$9\times10^{13}$ GeV, with the tensor to scalar ratio $r$
predicted to be $\gtrsim0.02$. Proton lifetime estimates for decays mediated by
the superheavy gauge bosons are also provided.}

\begin{document} \maketitle \flushbottom

\section{Introduction} \label{intro}
The observed quantization of electric charge is elegantly explained by invoking
the presence of magnetic monopoles, as shown by Dirac more than eighty years ago
\cite{Dirac:1931kp}. Contemporary unified theories with electric charge quantization based
on groups such as $SU(4) \times SU(2)_L \times SU(2)_R$ \cite{Pati:1974yy}, 
$SU(5)$ \cite{Georgi:1974sy}, $SO(10)$ or $E_6$,
predict the existence of topologically stable magnetic monopoles \cite{'tHooft:1974qc}, and one
expects that these monopoles are produced in the early universe.

Despite the presence of fractionally
charged quarks, the lightest $SU(5)$ monopole carries a single unit of Dirac
magnetic charge. 
This comes about because the unbroken gauge symmetry $SU(3)_c
\times U(1)_{em}$ share a $Z_3$ symmetry \cite{Daniel:1979yz}. 
The $SU(5)$ monopole ends up carrying some
color magnetic flux that is screened due to color confinement. The $SU(5)$
monopoles are superheavy with a mass about an order of magnitude larger than
$M_{{\rm GUT}} \sim2\times 10^{16}$ GeV.

In non-supersymmetric GUTs such as $SO(10)$ broken to the Standard Model (SM) via
$G_{422} = SU(4)_c \times SU(2)_L \times SU(2)_R$, there appears a new scenario
for monopole charges and masses. The $SO(10)$ breaking
to $G_{422}$ yields, just as in $SU(5)$, a
superheavy monopole with a single unit of Dirac magnetic charge  \cite{Lazarides:1980cc}. The
subsequent breaking of $G_{422}$ at some intermediate mass scale $M_I$ yields monopoles
that carry two units of Dirac charge and mass that can be a few orders of
magnitude smaller than the mass of the SU(5) monopole \cite{Lazarides:1980cc}.

It was argued a long time ago by Lazarides and Shafi \cite{Lazarides:1984pq} that within the
framework of GUT inflation driven by a gauge singlet scalar inflaton field
\cite{Shafi:1983bd},
these somewhat lighter monopoles may not be entirely inflated 
away.\footnote{For an earlier discussion of this with cosmic strings, see 
ref. \cite{Shafi:1984tt}.} The
superheavy monopoles produced during the first stage of symmetry breaking
experience at least the 50-60 e-foldings of observable inflation. The somewhat lighter
monopoles, produced during the intermediate symmetry breaking with mass
determined by $M_I$ and comparable to the Hubble constant $H$ during inflation, may
undergo a significantly reduced number of e-foldings. Therefore, there arises
the exciting possibility that these monopoles, lighter than $M_{{\rm GUT}}$, may be
present in our galaxy at an observable number density, comparable to or a few
orders of magnitude below the Parker bound \cite{Parker:1970xv}. 

In recent years the WMAP \cite{Hinshaw:2012aka} and Planck
\cite{Ade:2015xua,Ade:2015lrj} satellite experiments have provided a fairly
accurate determination of the scalar spectral index $n_s$ and an upper bound for
the tensor to scalar ratio $r \lesssim 0.1$. In the framework of GUT inflation
driven by a gauge singlet scalar field, one finds that for $n_s\ge 0.96$, the
energy scale during inflation is of order $10^{16}$ GeV
\cite{Shafi:2006cs,Rehman:2008qs}. 

In this brief report we calculate the range of energy scales during non-SUSY GUT
inflation such that $n_s$ and $r$ are compatible with the Planck 2015 constraints
\cite{Ade:2015xua}. This determines the magnitude
of $H$ which, in turn, provides an estimate for the range for $M_I$ that is
compatible with an observable flux of primordial magnetic monopoles.
These monopoles with mass $\sim 10^{14}$ GeV do not necessarily
catalyze nucleon decay with a strong interaction rate, and they should be
accessible to current and future large scale detectors. Estimates for the proton
lifetime are also provided.

\section{Inflation with Coleman-Weinberg potential} \label{cw}

The first new inflation models \cite{Linde:1981mu} 
were proposed in the early eighties immediately after Guth's seminal paper
\cite{Guth:1980zm}. They were based on $SU(5)$ GUT, with symmetry breaking due
to the Coleman-Weinberg mechanism \cite{Coleman:1973jx} occurring in the adjoint
Higgs field. However, it was shown in ref. \cite{Shafi:1983bd}
that obtaining sufficiently small density perturbations was
only possible if the scalar field $\phi$ is a gauge singlet. In these
Shafi-Vilenkin type models the field $\phi$ has a quartic potential at tree level, and
taking into account radiative corrections the potential becomes
(omitting terms that don't play an essential role) \cite{Albrecht:1984qt,Linde:2005ht}:
\begin{equation}
V=\frac{\lambda}{4}\phi^4-\frac12\beta^2\phi^2\chi^2+\frac
a4\chi^4+A\phi^4\left[\ln\left( \frac{\phi}{M}\right)+C\right]+V_0\,,
\end{equation}
where $\chi$ represents the field breaking the GUT group, 
$A\sim(1/16\pi^2)\beta^4$, $M$ and $C$ are normalization parameters and
$V_0\equiv V(\phi=0)$ is the vacuum energy density at the origin.
The $\chi$ field can be replaced by its vacuum expectation value (VEV) $\langle\chi\rangle=(\beta/\sqrt
a)\phi$. The parameter $C$ can be fixed by taking $M$ to be the $\phi$ VEV at the
minimum. Requiring $V(\phi=M)=0$ fixes $V_0 = A M^4 /4$. Also taking
$\lambda\ll\beta^4/a$, the effective potential
takes the standard form \cite{Albrecht:1984qt,Linde:2005ht}:
\begin{equation} \label{potpot}
                  V(\phi)= A \phi^4 \left[\ln\left( \frac{\phi}{M}\right)
-\frac{1}{4}\right] + \frac{A M^4}{4}\,.
\end{equation}
The inflationary predictions of this potential were recently analyzed in
ref. \cite{Shafi:2006cs} (see also refs.
\cite{Smith:2008pf,Rehman:2008qs}).

The magnitude of $A$ and the inflationary parameters can be calculated using the
standard slow-roll expressions. The slow-roll parameters may be defined as
(see ref. \cite{Lyth:2009zz}
for a review and references):
\begin{equation}
\epsilon =\frac{1}{2}\left( \frac{V^{\prime} }{V}\right) ^{2}\,, \quad
\eta = \frac{V^{\prime \prime} }{V}  \,, \quad
\xi ^{2} = \frac{V^{\prime} V^{\prime \prime\prime} }{V^{2}}\,.
\end{equation}
Here and below we use units $m_P=2.44\times10^{18}\rm{~GeV}=1$, 
and primes denote derivatives with respect
to the inflaton field $\phi$.
The spectral index
$n_s$, the tensor to scalar ratio
$r$ and the running of the spectral index
$\alpha\equiv\mathrm{d} n_s/\mathrm{d} \ln k$ are given in the slow-roll
approximation by
\begin{equation}
n_s = 1 - 6 \epsilon + 2 \eta \,,\quad
r = 16 \epsilon \,,\quad
\alpha = 16 \epsilon \eta - 24 \epsilon^2 - 2 \xi^2\,.
\end{equation}

The amplitude of the curvature perturbation $\Delta_\mathcal{R}$ is given by
\begin{equation} \label{perturb}
\Delta_\mathcal{R}=\frac{1}{2\sqrt{3}\pi}\frac{V^{3/2}}{|V^{\prime}|}\,,
\end{equation}
which should satisfy $\Delta_\mathcal{R}^2\approx 2.4\times10^{-9}$
from the Planck measurement \cite{Ade:2015xua} with the pivot scale chosen at
$k_* = 0.002$ Mpc$^{-1}$.\footnote{Note that while the Planck collaboration
otherwise uses a pivot scale corresponding to 0.05 Mpc$^{-1}$, they present their results on $r$ 
using $k_* = 0.002$ Mpc$^{-1}$. To facilitate comparason with the Planck results
we also take $k_* = 0.002$ Mpc$^{-1}$.}

The number of e-folds is given by
\begin{equation} \label{efold1}
N_*=\int^{\phi_*}_{\phi_e}\frac{V\rm{d}\phi}{V^{\prime}}\,, \end{equation}
where the subscript ``$_*$'' denotes quantities when the scale corresponding to
$k_*$ exited the horizon, 
and $\phi_e$ is the inflaton value at the end of inflation, which we
estimate by $\epsilon(\phi_e) = 1$. 

For $V_0^{1/4}\gtrsim2\times10^{16}$ GeV, observable inflation occurs close to
the minimum where the potential is effectively quadratic ($V\simeq2AM^2 \chi^2$,
where $\chi=\phi-M$ denotes the
deviation of the field from the minimum).
The inflationary predictions are thus approximately given by
\begin{equation}
\label{quadratic}
n_s = 1-2/N\,,\quad r =8/N\,, \quad \alpha =-2/N^2\,.
\end{equation}

For $V_0^{1/4}\lesssim10^{16}$ GeV, assuming inflation takes place with inflaton
values below $M$, the inflationary parameters are
similar to those for new inflation models with $V=V_0[1-(\phi/\mu)^4]$:
$n_s\simeq1-(3/N)$, $r$ small, and
$\alpha\simeq -3/N^2$. 

Note that in the context of non-supersymmetric GUTs, $V_0^{1/4}$ is related to
the unification scale $M_U$, and is typically a factor of $\sim\sqrt{4\pi}$
smaller than the superheavy gauge boson masses due to the loop factor in the
Coleman-Weinberg potential. The allowed range of $V_0^{1/4}$ (and hence of
$M_U$) can be calculated by comparing the $n_s$ and $r$ values with the  Planck
results \cite{Ade:2015xua}. Since the resulting constraints depend
sensitively on the number of e-folds $N$, instead of fixing $N$ to a fiducial
value, we calculate it using
\begin{equation} \label{efolds}
N_*\approx64.7+\frac12\ln\frac{\rho_*}{m^4_P}-\frac{1}{3(1+\omega_r)}\ln\frac{\rho_e}{m^4_P}
+\left(\frac{1}{3(1+\omega_r)}-\frac14\right)\ln\frac{\rho_r}{m^4_P}\,.
\end{equation}
Here $\rho_{e}=(3/2)V(\phi_{e})$ is the
energy density at the end of inflation, $\rho_r$ is the energy density at the
end of reheating and $\omega_r$ is the equation of state parameter during
reheating, which we take to be constant. For a derivation of \eq{efolds} see
e.g. ref. \cite{Liddle:2003as}.

\begin{figure}[!tbh]
\begin{center}
\scalebox{0.54}{\includegraphics{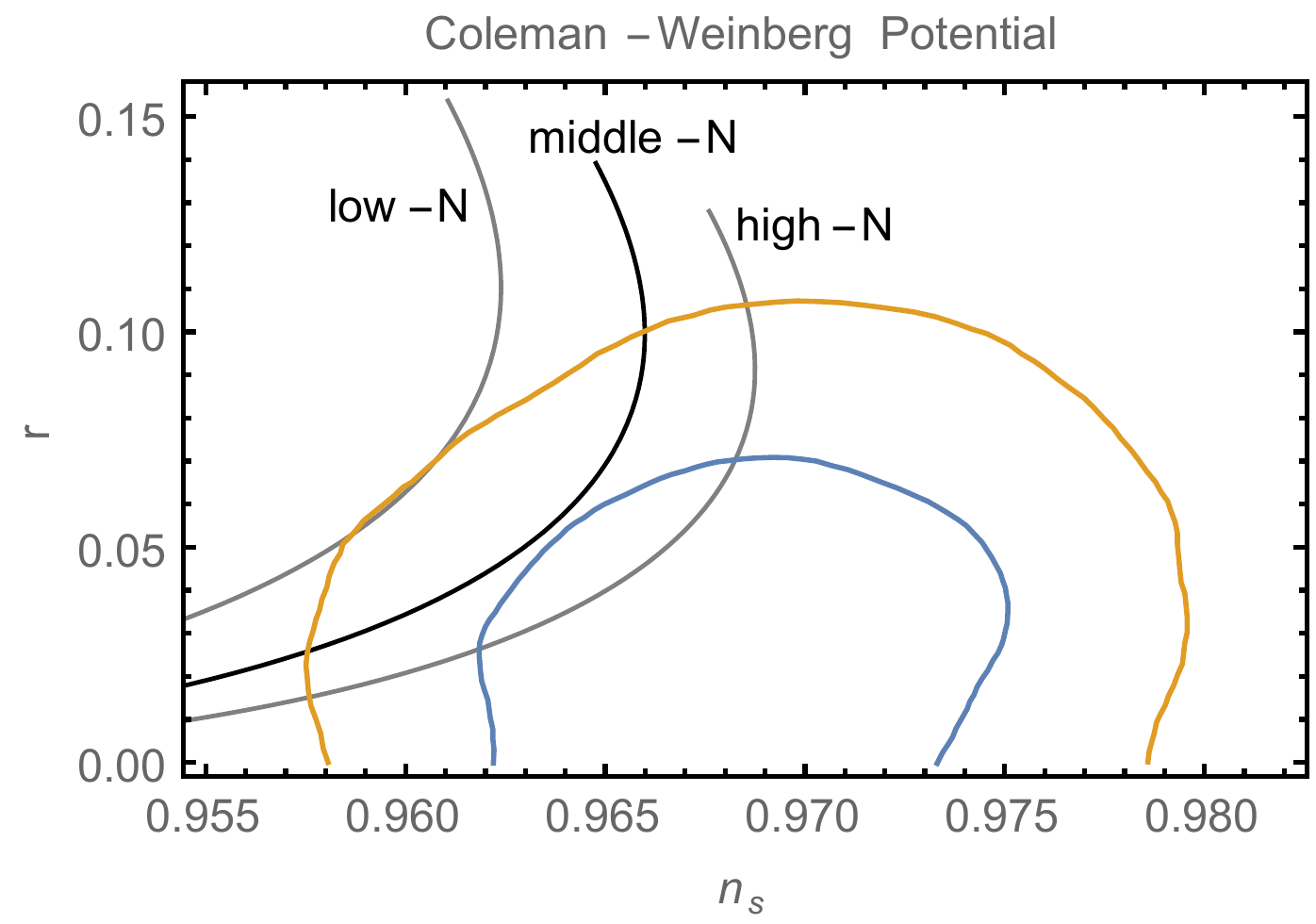}}
\scalebox{0.54}{\includegraphics{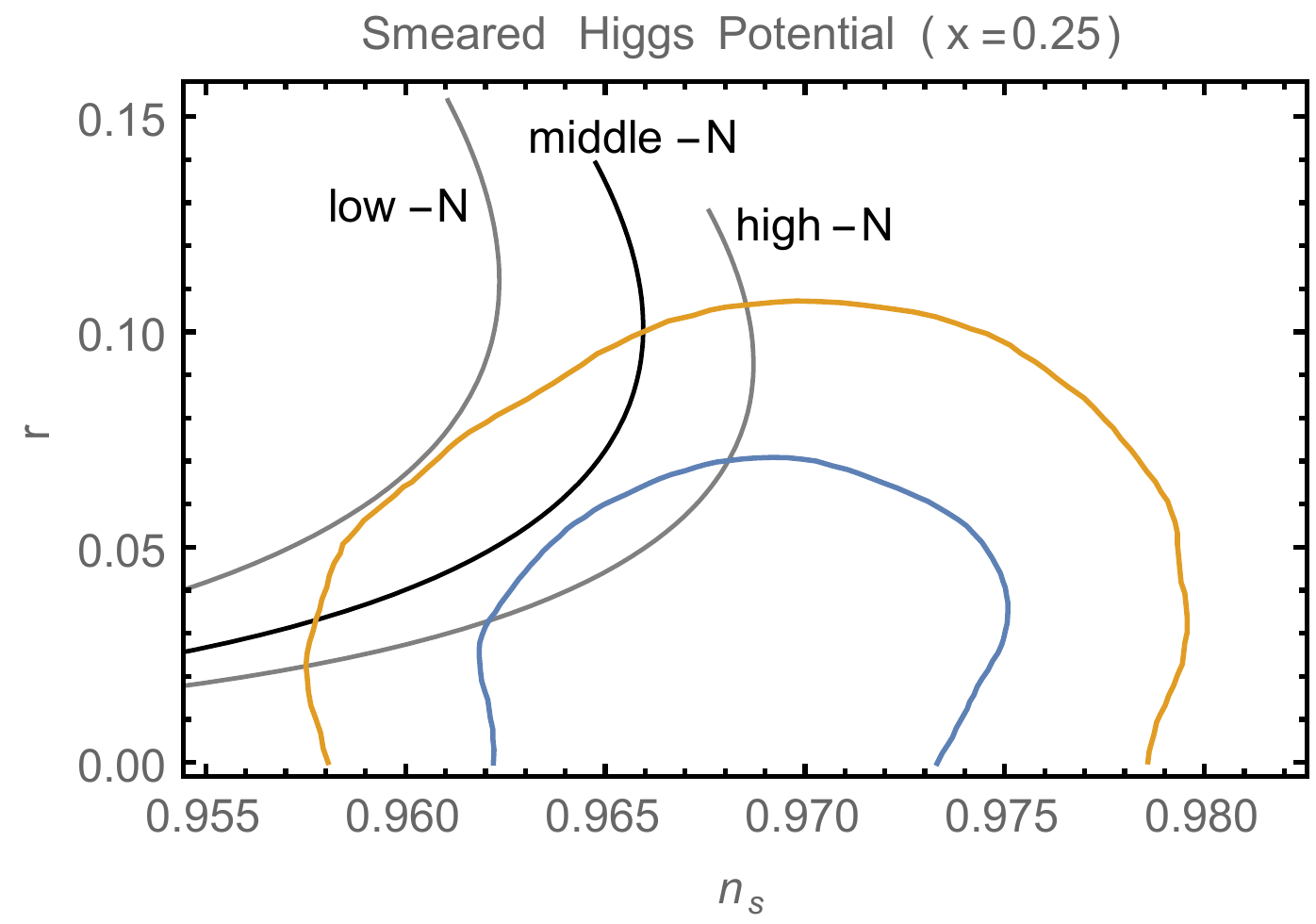}}\\ \vspace{0.3cm}
\scalebox{0.54}{\includegraphics{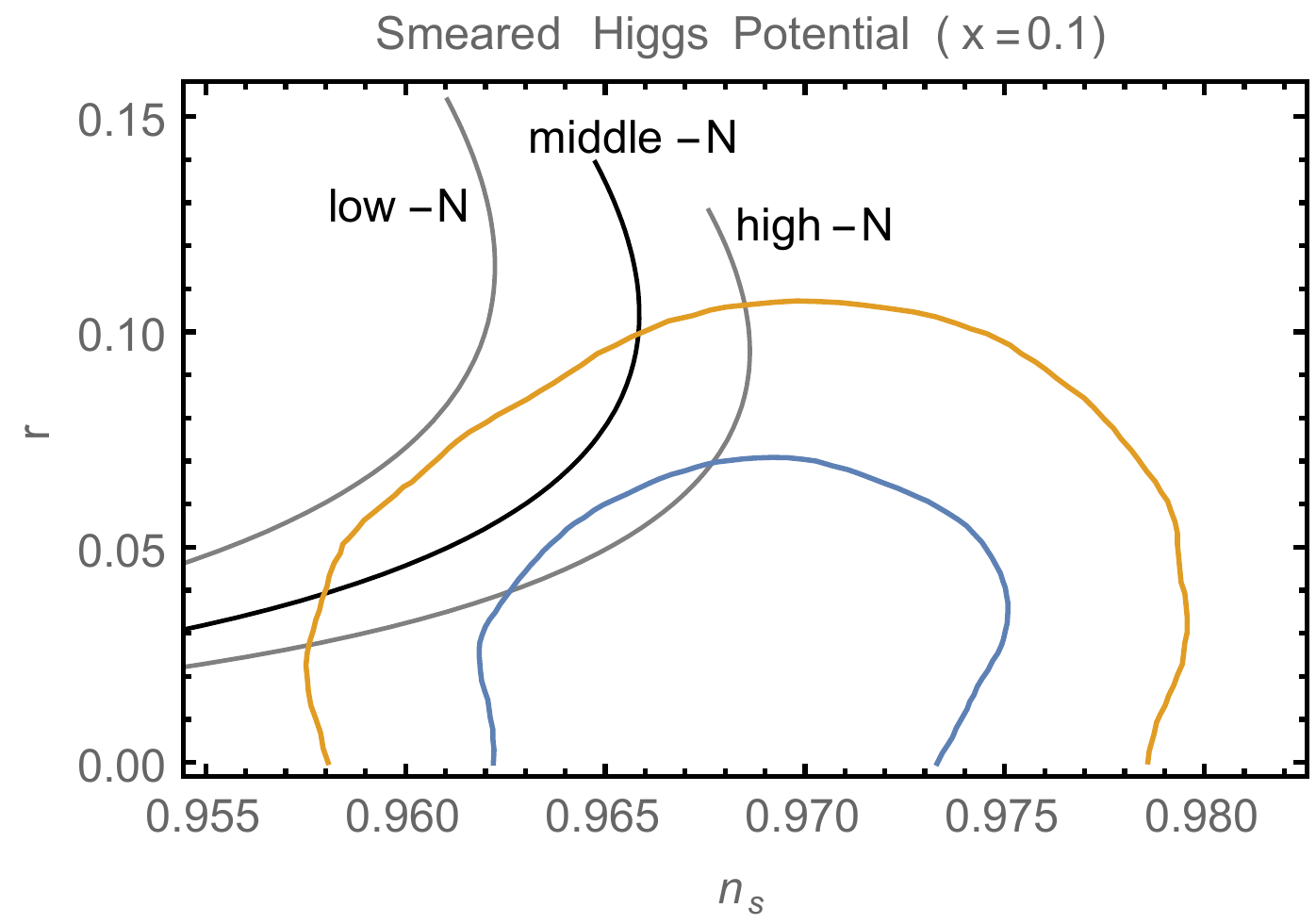}}
\scalebox{0.54}{\includegraphics{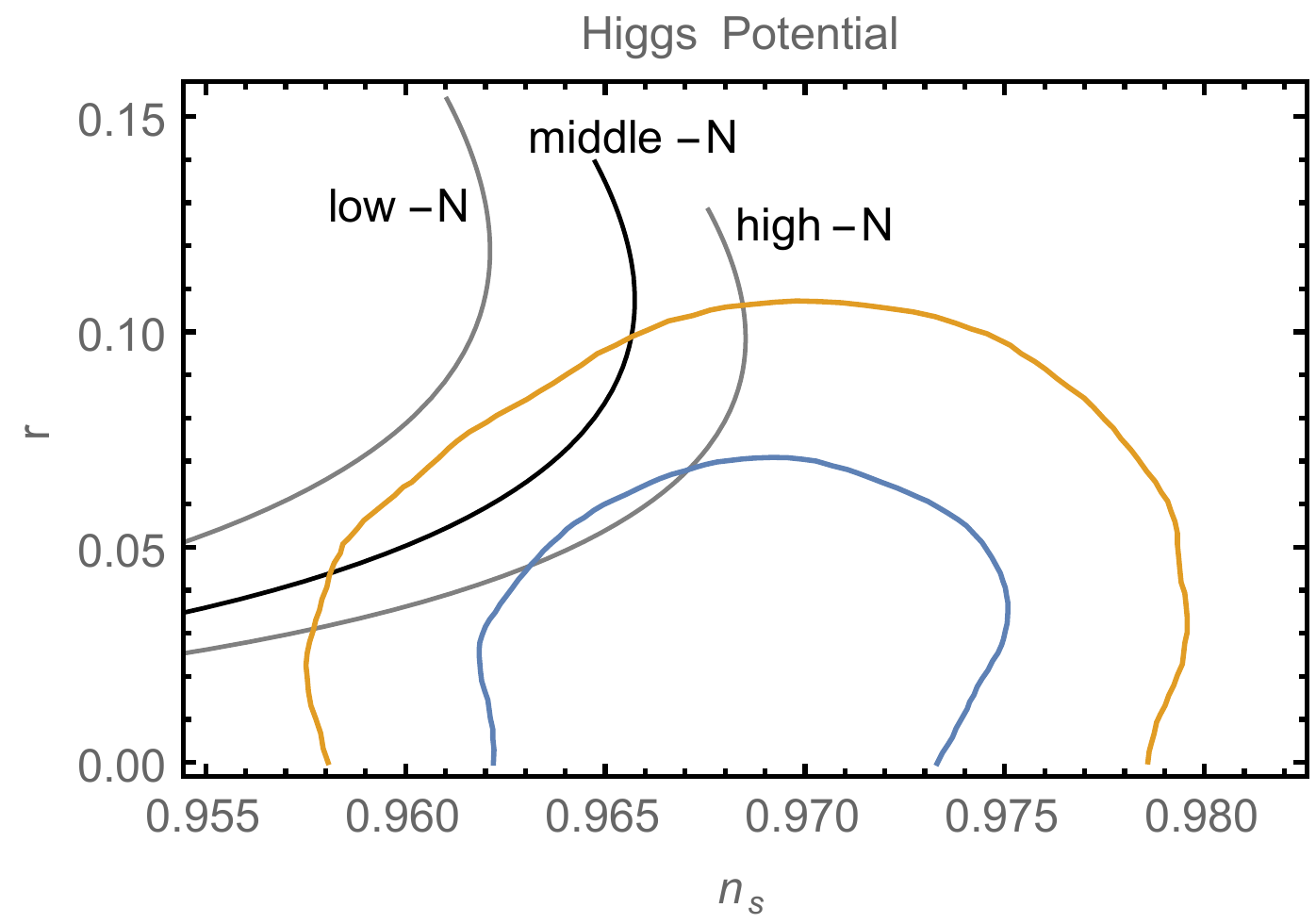}}
\end{center}\vspace{-0.5cm}
\caption{$n_s$ vs. $r$ curves 
along with the  68\% and 95\% confidence level contours  
 given by the Planck collaboration (Planck TT+lowP+BKP+lensing+ext) \cite{Ade:2015xua}.}
  \label{inf_figure}
\end{figure}

To represent a plausible range of $N$, we consider three cases: In the high-$N$
case $\omega_{r}$ is taken to be 1/3, which is equivalent to assuming instant
reheating.  In the middle-$N$ case we take $\omega_{r}=0$ and the reheat temperature
$T_r=10^9$ GeV, calculating $\rho_r$ using the SM value for the
number of relativistic degrees of freedom ($g_*=106.75$). 
In the low-$N$ case we take $T_r=100$ GeV (again with $\omega_{r}=0$).\footnote{$T_r$ as low
as 10 MeV is consistent with big bang nucleosynthesis, however it is difficult
to explain how baryogenesis could occur at such low temperatures.} The $n_s$
vs. $r$ curve for each case is shown in Figure \ref{inf_figure} along  with the contours (at
the confidence levels of 68\% and 95\%) given by the Planck collaboration
(Planck TT+lowP+BKP+lensing+ext) \cite{Ade:2015xua}. Numerical
results for selected values of $V_0$ and the middle-$N$ case are displayed in
Table \ref{inf_tab}.  

\begin{table}[!tbh]
\begin{center}
\begin{tabular}{cccccccccccc}
\hline \hline
 {\small $V_0^{1/4}$} &{\small  $V(\phi_0)^{1/4}$} & 
$\log(A)$ & $m$ & $M$ & $\phi_*$ &
$\phi_e$ &  $N_*$ & $n_s$  & $r$ & $-\alpha$     \\
{\scriptsize$10^{16}$GeV} & {\scriptsize$10^{16}$GeV} &
&{\scriptsize$10^{13}$GeV}&{\scriptsize$m_P$} &{\scriptsize$m_P$} & {\scriptsize$m_P$}& & & &
{\scriptsize$10^{-4}$} \\
\hline \hline
\multicolumn{11}{c}{Coleman-Weinberg potential}\\\hline
$1.25 $ &$ 1.23 $ & -13.3 && 15.2 & 5.13 & 13.9 & 55.1 & 0.955 & 0.018 & 6.20\\\hline
$1.5  $& $1.44  $& -13.4 &&19.9 & 8.64 & 18.5 & 55.3 & 0.960 & 0.034 & 5.74\\\hline
$1.75  $& $1.61  $& -13.6 && 26.2 & 14.0 & 24.8 & 55.5 &
0.963 & 0.054 & 5.74\\\hline
$2.0  $&$ 1.73  $& -13.9 && 34.3 & 21.4 & 33.0 & 55.6 & 0.965 & 0.072 & 5.87\\\hline
$3.0 $& $1.94  $& -14.7 && 83.1 & 68.9 & 81.7 & 55.9 & 0.966 & 0.112 & 6.12\\\hline
$6.0  $& $2.03  $& -16.0 && 356 & 341 & 354 & 56.0 & 0.965 & 0.135 & 6.23\\\hline
\hline
\multicolumn{11}{c}{Smeared Higgs potential ($x=0.25$)}\\\hline
1.25 & 1.23 & -13.6 & 0.88 & 12.6 & 2.31 & 11.3 & 55.1 & 0.95 & 0.019 & 4.59\\\hline
1.5 & 1.45 & -13.7 & 1.00 & 16.0 & 4.66 & 14.7 & 55.3 & 0.958 & 0.035 & 5.23\\\hline
1.75 & 1.62 & -13.9 & 1.04 & 20.9 & 8.60 & 19.5 & 55.5 & 0.963 & 0.055 & 5.61\\\hline
2.0 & 1.74 & -14.1 & 1.04 & 27.3 & 14.3 & 26.0 & 55.7 & 0.965 & 0.074 & 5.85\\\hline
3.0 & 1.94 & -14.9 & 0.97 & 65.9 & 51.7 & 64.5 & 55.9 & 0.966 & 0.114 & 6.13\\\hline
6.0 & 2.03 & -16.2 & 0.91 & 281 & 267 & 280 & 56.0 & 0.965 & 0.135 & 6.23\\\hline
\hline
\multicolumn{11}{c}{Smeared Higgs potential ($x=0.1$)}\\\hline
1.25  &  1.24  &  -13.9  &  0.98  &  12.5  &  2.00  &  11.2  &  55.2  &  0.946  &  0.019  &  3.71\\\hline
1.5  &  1.45  &  -14.0  &  1.13  &  15.6  &  4.05  &  14.2  &  55.4  &  0.957  &  0.036  &  4.78\\\hline
1.75  &  1.63  &  -14.2  &  1.19  &  20.0  &  7.59  &  18.6  &  55.5  &  0.962  &  0.056  &  5.46\\\hline
2.0  &  1.75  &  -14.4  &  1.20  &  26.0  &  12.9  &  24.6  &  55.7  &  0.965  &
0.075  &  5.80\\\hline
3.0  &  1.95  &  -15.2  &  1.13  &  62.1  &  47.9  &  60.7  &  55.9  &  0.966  &  0.115  &  6.14\\\hline
6.0  &  2.03  &  -16.5  &  1.06  &  264  &  249  &  263  &  56.0  &  0.965  &  0.135  &  6.23\\\hline
\hline
\multicolumn{11}{c}{Higgs potential}\\\hline
$1.25 $ & $1.24 $ && 1.03&12.5 & 1.85 & 11.2 & 55.2 & 0.943 & 0.019 & 3.16\\\hline
$1.5  $& $1.46  $& &1.20&15.4 & 3.72 & 14.0 & 55.4 & 0.955 & 0.036 & 4.45\\\hline
$1.75 $& $1.63 $ & &1.29&19.5 & 6.98 & 18.1 & 55.6 & 0.962 & 0.057 & 5.32\\\hline
$2.0  $& $1.76  $& &1.31&25.1 & 11.9 & 23.7 & 55.7 & 0.964 & 0.077 & 5.77\\\hline
$3.0  $& $1.95 $& &1.24&59.6 & 45.2 & 58.2 & 55.9 & 0.966 & 0.116 & 6.14\\\hline
$6.0  $& $2.03 $& &1.17&252 & 237 & 251 & 56.0 & 0.965 & 0.136 & 6.23\\\hline
\hline
\end{tabular}
\end{center}
\caption{Parameter values for the middle-$N$ case. 
} \label{inf_tab}
\end{table}

\section{Inflation with smeared Higgs potential} \label{higgs}
A generalization of the model considered in \sektion{cw} is to take a tree-level Higgs
potential $V=-(1/2)m^2\phi^2+(1/4)\lambda\phi^4$. The effective potential
\eq{potpot} then becomes the
sum of a Higgs potential and the Coleman-Weinberg potential
considered in \sektion{cw}
\cite{Rehman:2010es}:
\begin{equation} \label{potpot2}
V(\phi)=\left(\frac{m^2
M^2}{4}\right)\left[1-\left(\frac{\phi}{M}\right)^2\right]^2+
 A \phi^4 \left[\ln\left( \frac{\phi}{M}\right)
-\frac{1}{4}\right] + \frac{A M^4}{4}\,.
\end{equation}
Following ref. \cite{Rehman:2010es}, we will call $V(\phi)$ in \eq{potpot2} the
smeared Higgs potential. Depending on the value of $m$, the inflationary predictions
for this potential interpolate between the predictions for the tree-level Higgs
potential and for the Coleman-Weinberg potential \cite{Rehman:2010es}. 

The tree-level Higgs potential has been analyzed in several papers, see e.g. refs.
\cite{Vilenkin:1994pv,Smith:2008pf,Rehman:2008qs}.
The inflationary predictions are similar to the predictions for the
Coleman-Weinberg potential. For $V_0^{1/4}\gtrsim2\times10^{16}$ GeV, observable inflation occurs close to
the minimum where the potential is effectively quadratic ($V\simeq m^2 \chi^2$,
where $\chi=\phi-M$ denotes the
deviation of the field from the minimum).
The inflationary predictions are thus approximately given by \eq{quadratic}.
For $V_0^{1/4}\lesssim10^{16}$ GeV, assuming inflation takes place with inflaton
values below $M$, a red spectrum is predicted with $n_s\simeq1-8/M^2$. Compared
with the Coleman-Weinberg potential, the Higgs potential predicts 
higher values of $r$ for the same $n_s$ values.  

We represent the ``smearing'' of the Higgs potential by radiative corrections
with a smearing parameter $x$, where $AM^4/4=xV_0$ and $m^2M^2/4=(1-x)V_0$. With
this definition $x\to0$ and $x\to1$ corresponds to the Higgs and
Coleman-Weinberg potentials, respectively.

At the end of inflation, the GUT symmetry breaking fields have a VEV
$\langle\chi\rangle\sim(\beta/\sqrt a)M$. Taking $a\sim g^2$, where $g$ is the
gauge coupling, the unification scale is given by
\begin{equation} \label{MU}
M_U\sim\beta M\sim\sqrt{4\pi}(xV_0)^{1/4}\,.
\end{equation}
The $n_s$ vs. $r$ curves for tree-level and smeared Higgs potentials (for $x=0.25$
and 0.1) are shown in Figure \ref{inf_figure}. Numerical
results for selected values of $V_0$ and the middle-$N$ case are displayed in
Table \ref{inf_tab}.

\section{Magnetic monopoles and proton decay in non-supersymmetric GUTs}
\label{monopole}

The models discussed in sections \ref{cw} and \ref{higgs} can be realized within
the framework of non-supersymmetric GUTs such as those based on $SO(10)$ as well
as $SU(5)$, as discussed in ref. \cite{Lazarides:1984pq}.  The breaking of
$SO(10)$ to the SM can proceed, for example, via the intermediate
group $G_{422} = SU(4)_c \times SU(2)_L \times SU(2)_R$ \cite{Pati:1974yy}.
The monopoles associated with the breaking at scale $M_U$ of the GUT group to the
intermediate group are inflated away. However, the breaking of
the intermediate group to the SM gauge symmetry at the intermediate scale $M_I$
yields monopoles (doubly charged in the case of $G_{422}$
\cite{Lazarides:1980cc}), whose mass is an order
of magnitude larger than $M_I$.  These may be present in our galaxy at a flux
level that depends on the values of $V_0$ and $M_I$. Below we will estimate the
$M_I$ scale that corresponds to an observable
flux level following the arguments in Ref. \cite{Lazarides:1984pq}.

First let us consider the potential for the $\chi$ fields breaking the GUT group 
to the intermediate group, at scale $M_U$. The potential involves a thermal term
\begin{equation}
V\supset\frac12\sigma_{\chi}T_H^2\chi^2-\frac12\beta^2\phi^2\chi^2+\frac
a4\chi^4\,,
\end{equation}
where $T_H\equiv H/2\pi$ is the Hawking temperature and the coefficient
$\sigma_{\chi}\sim1$. Thus symmetry breaking occurs when
$\beta^2\phi^2\gtrsim(H/2\pi)^2$, and topological structures are ``frozen'' in
soon afterwards \cite{Kibble:1976sj}. It can be easily checked that this happens
much earlier than the horizon exit of cosmological scales, so as mentioned above
any such topological structures are inflated away.

Let $X$ denote the fields whose VEV breaks the intermediate group to the SM at the scale $M_I$.
This breaking occurs due to the coupling $-(1/2)c^2\phi^2 X^2$, where $c\sim M_I/M$.
Thus, symmetry breaking occurs and subsequently the monopoles are ``frozen'' in when
\begin{equation} \label{phix}
\phi\sim\phi_x\equiv\frac{H_x}{2\pi}\frac{M}{M_I}\,,
\end{equation}
where $H_x=(V(\phi_x)/3)^{1/2}$ is the Hubble constant when $\phi=\phi_x$.

If $M$ is small compared to the Planck scale, the inflaton $\phi$ is essentially
constant until almost the end of inflation, rolling quickly only within the last $H^{-1}$
\cite{Linde:2005ht}. This means for substantial dilution of the monopoles, $\phi_x$ should
be very close to $\phi_*$. On the other hand, if $M$ is large compared to the Planck scale
both $\phi_*$ and $\phi_x$ values will be close to $M$. Since the Planck constraint on
$n_s$ is only satisfied for this latter case, we have $\phi_x\approx M$ and
therefore $M_I\sim H_x /2\pi\sim10^{13}$ GeV. 

To be more specific, let's consider how much dilution of the monopoles is necessary.
$M_I\sim10^{13}$ GeV corresponds to monopole masses of order $M_M\sim10^{14}$ GeV.
For these intermediate mass monopoles the MACRO experiment has put an upper
bound on the flux of $2.8\times10^{-16}\ {\rm cm}^{-2}$ s$^{-1}$ sr$^{-1}$
\cite{Ambrosio:2002qq}.  For monopole mass $\sim10^{14}$ GeV, this bound corresponds
to a monopole number per comoving volume of $Y_M\equiv
n_M/s\lesssim10^{-27}$ \cite{Kolb:1990vq}. There is also a stronger but indirect
bound on the flux of ($M_M/10^{17}$ GeV)$10^{-16}{\rm cm}^{-2}$ s$^{-1}$ sr$^{-1}$
obtained by considering the evolution of the seed Galactic magnetic field
\cite{Adams:1993fj}.

Direct search bounds stronger than the MACRO bound were obtained in refs.
\cite{Ueno:2012md}, but these apply to monopoles that catalyze
nucleon decay through the Callan-Rubakov process
\cite{Callan:1982au}. There are even more stringent indirect
bounds from compact astrophysical objects capturing monopoles
\cite{Agashe:2014kda}.  However, the monopoles produced during the intermediate
symmetry breaking stage do not necessarily catalyze nucleon decay (at least, not
with a strong interaction rate) \cite{Dawson:1982sc}.
This improves the chances of directly observing such monopoles in the future,
since the bounds from compact astrophysical objects are avoided.

At production, the monopole number density $n_M$ is of order $H_x^3$
\cite{Kibble:1976sj,Lazarides:1984pq}, which gets diluted to $H_x^3e^{-3N_x}$, where $N_x$ is
the number of $e$-folds after $\phi=\phi_x$. Using 
\begin{equation} \label{YM}
Y_M\sim\frac{H_x^3e^{-3N_x}}{s}
\,,
\end{equation}
where $s=(2\pi^2g_S/45)T_r^3$, we find that sufficient dilution requires
$N_x\gtrsim\ln (H_x/T_r)+20$. Thus, for $T_r\sim10^9$ GeV, $N_x\gtrsim30$ yields a monopole flux
close to the observable level. 

\begin{figure}[!t]
\begin{center}
\scalebox{0.54}{\includegraphics{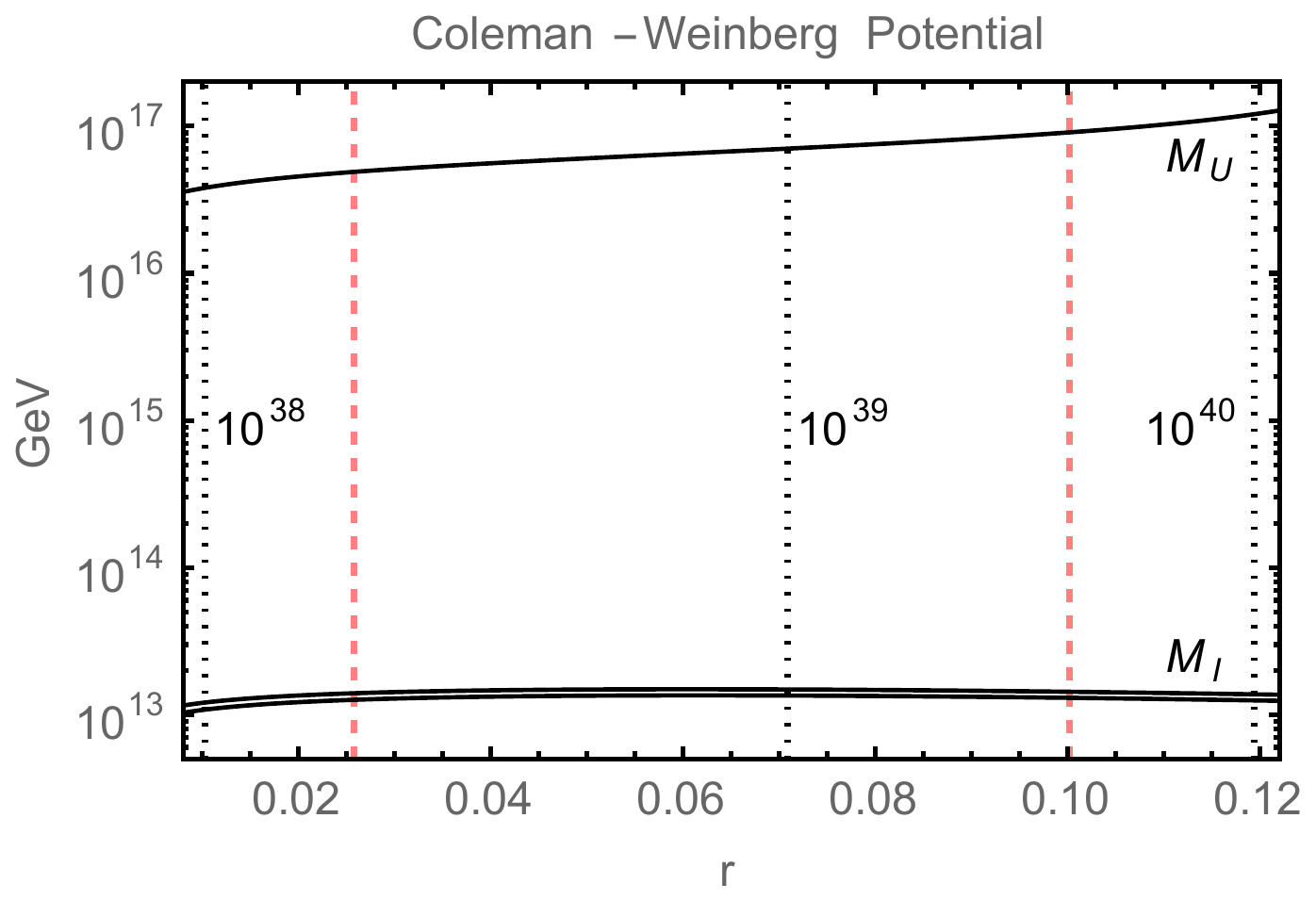}}
\scalebox{0.54}{\includegraphics{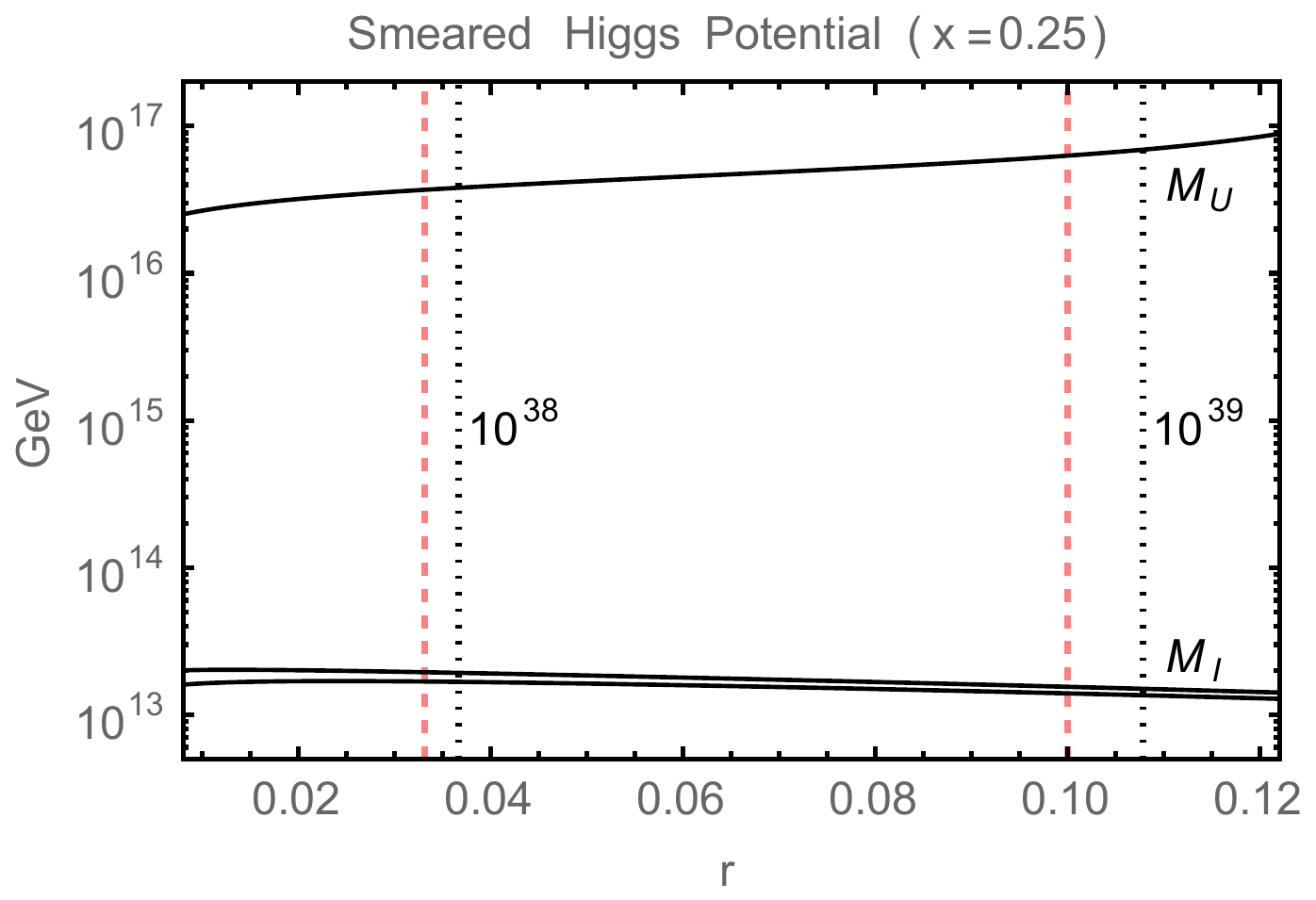}}\\ \vspace{0.3cm}
\scalebox{0.54}{\includegraphics{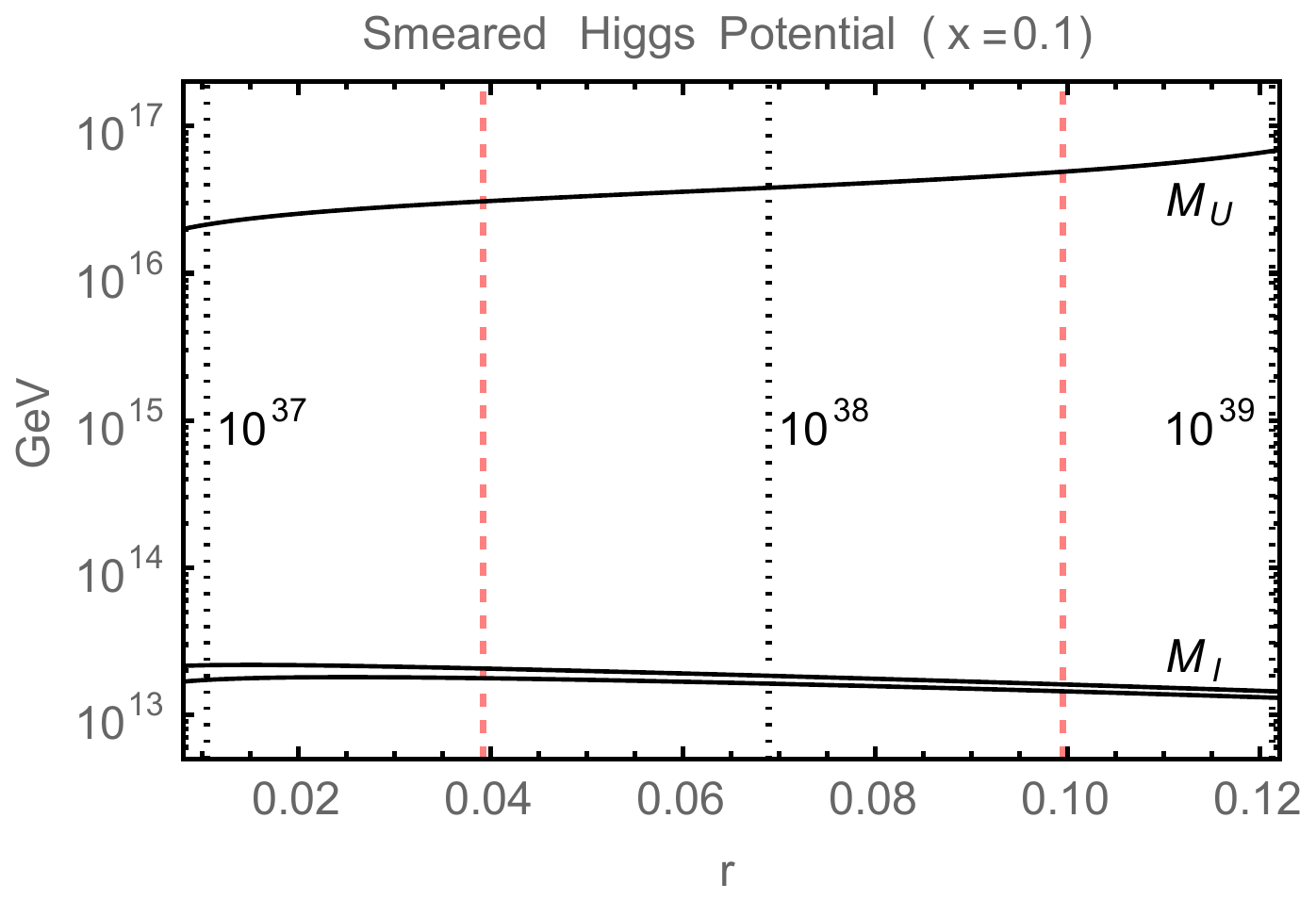}}
\scalebox{0.54}{\includegraphics{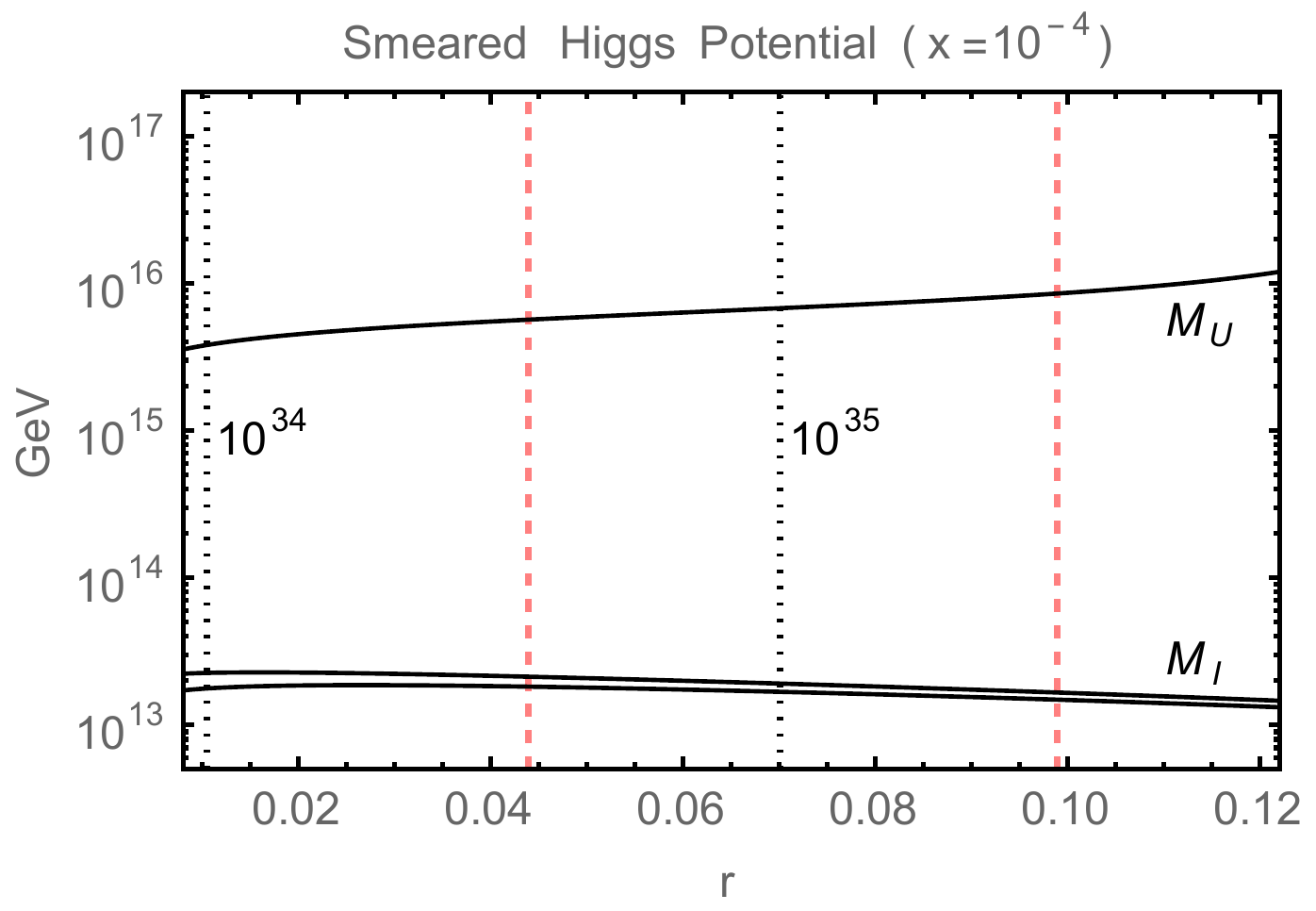}}
\end{center}\vspace{-0.5cm}
\caption{$M_U$ values and the range of $M_I$ values (corresponding to a thin band around
$10^{13}$ GeV) that give an observable monopole flux (between $2.8\times10^{-16}$
and $10^{-24}\ {\rm cm}^{-2}$ s$^{-1}$ sr$^{-1}$)  are
shown as a function of $r$ for the middle-N case. The  dotted vertical lines
show the proton lifetime in years. The region between the  dashed vertical lines
correspond to $n_s$ and $r$ values within the 95\% confidence level contours  
 given by the Planck collaboration (Planck TT+lowP+BKP+lensing+ext) \cite{Ade:2015xua}.
}
  \label{monfigure}
\end{figure}

Using \eq{YM}, we calculate $\phi_x$, $H_x$ and $N_x$ values, denoted with
subscripts ``$_+$'' and ``$_-$'', corresponding respectively to the flux levels
$2.8\times10^{-16}\ {\rm cm}^{-2}$ s$^{-1}$ sr$^{-1}$ which is the MACRO
bound and $10^{-24}\ {\rm cm}^{-2}$ s$^{-1}$ sr$^{-1}$ which we take as a rough
threshold for observability.  Then using \eq{phix}, we calculate the
corresponding $M_I$ values, which are shown in Figure
\ref{monfigure} and Table \ref{montable}.  For $M_I\gtrsim M_{-}$, the monopoles are too diluted to be
observable, whereas $M_I\lesssim M_{+}$ is excluded from the bound on the flux.

\setlength{\tabcolsep}{0.5em}
\begin{table}[!t]
\begin{center}
\begin{tabular}{ccccccccccc}
\hline \hline
 {\small $V_0^{1/4}$} & $\phi_+$ & $\phi_-$  & $H_+$ & $H_-$ & $N_+$ & $N_-$ &
$M_+$ & $M_-$ & $M_U$ & $\tau_p$   \\
{\scriptsize$10^{16}$GeV} &{\scriptsize$m_P$} & {\scriptsize$m_P$} &
{\scriptsize$10^{13}$GeV} & {\scriptsize$10^{13}$GeV} & & &
{\scriptsize$10^{13}$GeV} & {\scriptsize$10^{13}$GeV} &
{\scriptsize$10^{16}$GeV} & {\scriptsize years}
\\
\hline \hline
\multicolumn{11}{c}{Coleman-Weinberg potential}\\\hline
1.25 & 6.79 & 6.25 & 3.38 & 3.45 & 29.9 & 36.4 & 1.20 & 1.33 & 4.43 &
$2\times10^{38}$\\\hline
1.5 & 10.8 & 10.1 & 4.46 & 4.62 & 30.2 & 36.7 & 1.31 & 1.44 & 5.32 &
$4\times10^{38}$\\\hline
1.75 & 16.6 & 15.8 & 5.37 & 5.63 & 30.3 & 36.9 & 1.35 & 1.49 & 6.20 &
$7\times10^{38}$\\\hline
2.0 & 24.3 & 23.4 & 6.00 & 6.37 & 30.5 & 37.0 & 1.35 & 1.48 & 7.09 &
$1\times10^{39}$\\\hline
3.0 & 72.4 & 71.4 & 6.96 & 7.54 & 30.6 & 37.2 & 1.27 & 1.40 & 10.6 &
$6\times10^{39}$\\\hline
6.0 & 345 & 344 & 7.31 & 8.02 & 30.7 & 37.2 & 1.20 & 1.32 & 21.3 &
$1\times10^{41}$\\\hline
\hline
\multicolumn{11}{c}{Smeared Higgs potential ($x=0.25$)}\\\hline
1.25 & 4.03 & 3.46 & 3.39 & 3.48 & 29.9 & 36.4 & 1.69 & 2.01 & 3.13 &
$5\times10^{37}$\\\hline
1.5 & 6.85 & 6.16 & 4.51 & 4.67 & 30.2 & 36.7 & 1.68 & 1.93 & 3.76 &
$1\times10^{38}$\\\hline
1.75 & 11.2 & 10.4 & 5.42 & 5.69 & 30.4 & 36.9 & 1.61 & 1.82 & 4.39 &
$2\times10^{38}$\\\hline
2.0 & 17.3 & 16.4 & 6.06 & 6.43 & 30.5 & 37.0 & 1.53 & 1.71 & 5.01 &
$3\times10^{38}$\\\hline
3.0 & 55.2 & 54.2 & 6.98 & 7.57 & 30.6 & 37.2 & 1.33 & 1.47 & 7.52 &
$2\times10^{39}$\\\hline
6.0 & 270 & 269 & 7.32 & 8.03 & 30.7 & 37.2 & 1.21 & 1.34 & 15.0 &
$2\times10^{40}$\\\hline
\hline
\multicolumn{11}{c}{Smeared Higgs potential ($x=0.1$)}\\\hline
1.25 & 3.77 & 3.18 & 3.39 & 3.48 & 29.9 & 36.4 & 1.79 & 2.17 & 2.49 &
$2\times10^{37}$\\\hline
1.5 & 6.28 & 5.57 & 4.52 & 4.70 & 30.2 & 36.7 & 1.78 & 2.09 & 2.99 & $4\times10^{37}$\\\hline
1.75 & 10.2 & 9.43 & 5.46 & 5.74 & 30.4 & 36.9 & 1.70 & 1.94 & 3.49 &
$7\times10^{37}$\\\hline
2.0 & 15.8 & 14.9 & 6.10 & 6.48 & 30.5 & 37.0 & 1.59 & 1.79 & 3.99 & $1\times10^{38}$\\\hline
3.0 & 51.4 & 50.4 & 7.00 & 7.60 & 30.6 & 37.2 & 1.35 & 1.49 & 5.98 & $6\times10^{38}$\\\hline
6.0 & 253 & 252 & 7.32 & 8.03 & 30.7 & 37.2 & 1.22 & 1.34 & 12.0 & $1\times10^{40}$\\\hline
\hline
\multicolumn{11}{c}{Smeared Higgs potential ($x=10^{-4}$)}\\\hline
1.25 & 3.65 & 3.05 & 3.39 & 3.48 & 29.9 & 36.4 & 1.85 & 2.28 & 0.44 &
$2\times10^{34}$\\\hline
1.5 & 5.97 & 5.26 & 4.53 & 4.71 & 30.2 & 36.7 & 1.85 & 2.19 & 0.53 &
$4\times10^{34}$\\\hline
1.75 & 9.66 & 8.84 & 5.48 & 5.77 & 30.4 & 36.9 & 1.76 & 2.02 & 0.62 &
$7\times10^{34}$\\\hline
2.0 & 14.9 & 14.0 & 6.13 & 6.52 & 30.5 & 37.0 & 1.64 & 1.86 & 0.71 & $1\times10^{35}$\\\hline
3.0 & 48.8 & 47.8 & 7.02 & 7.63 & 30.6 & 37.2 & 1.36 & 1.51 & 1.06 &
$6\times10^{35}$\\\hline
6.0 & 241 & 240 & 7.33 & 8.04 & 30.7 & 37.2 & 1.22 & 1.34 & 2.13 &
$1\times10^{37}$\\\hline
\hline
\end{tabular}
\end{center}
\caption{Parameter values for the middle-N case. 
} \label{montable}
\end{table}

Another key prediction of GUTs besides magnetic monopoles is proton decay. The 
proton mean life can be estimated as
\begin{equation} \label{taup}
\tau_p\sim\frac{M_U^4}{\alpha_G^2 m_{pr}^5}
\,,
\end{equation}
where $M_U$ is estimated using \eq{MU}, $m_{pr}$ is the proton mass, and
$\alpha_G\sim1/40$ is the GUT coupling constant. Using \eq{taup}, the
experimental bound $\tau_p(p\to e^+\pi^0)>8.2\times10^{33}$ years
\cite{Nishino:2009aa} corresponds to $M_U\gtrsim4\times10^{15}$ GeV, whereas a
realistically observable $\tau_p(p\to e^+\pi^0)=10^{35}$ years \cite{Abe:2011ts}
corresponds to $M_U\approx8\times10^{15}$ GeV.  Since the Planck constraint on
$n_s$ is only satisfied for $V_0^{1/4}\gtrsim10^{16}$ GeV, a glance at \eq{MU}
shows that proton decay is typically too slow to be observed in this class of
models, unless the smearing parameter $x$ is close to zero. Proton lifetime
estimates are displayed in  Figure
\ref{monfigure} and Table \ref{montable}.

\section{Conclusion} \label{}

In this paper we discussed a class of models where the gauge-singlet inflaton
has either a quartic or Higgs potential at tree level. The radiative corrections
due to couplings with GUT symmetry breaking fields modify the tree level
potential into a Coleman-Weinberg or smeared Higgs potential.  If the GUT
symmetry breaking to the SM proceeds via an intermediate group, the breaking of
the intermediate group to the SM gauge symmetry at intermediate scale $M_I$
yields monopoles whose mass is an order of magnitude larger than $M_I$. These
may be present in our galaxy at a flux level that depends on the values of $V_0$
(the vacuum energy density at the origin) and $M_I$.

For both Coleman-Weinberg and smeared Higgs potentials the Planck constraint
$n_s>0.955$ is only satisfied for $V_0^{1/4}\gtrsim10^{16}$ GeV, which implies
$r\gtrsim0.02$, a level which can be probed in this decade. Another consequence
of the Planck constraint is that an observable level of monopole flux can only
occur for $M_I\sim10^{13}$ GeV, with lower values excluded due to excessive
monopole flux. Thus, a smoking gun evidence for this class of models would be
the observation of monopoles with masses of order $10^{14}$ GeV, together with
the observation of a B-mode CMB polarization signal corresponding to $r\gtrsim0.02$.

The lower bound on $M_I$ poses a severe constraint for $SO(10)$ broken to SM via
$G_{422}$, since the typical values obtained from the RG analysis is $M_I\sim
10^{11}$ GeV and $M_U\sim10^{16}$ GeV \cite{Lee:1994vp}.  However, taking
threshold effects due to the Higgs sector into account, it is possible to achieve $M_I$ as high as $3\times10^{13}$ GeV with $M_U\approx4\times10^{15}$
GeV \cite{Lee:1994vp,Acampora:1994rh}.\footnote{Threshold effects from gauge
invariant higher dimensional operators can also significantly modify the
standard predictions for $M_I$ and $M_U$ \cite{Shafi:1983gz}.}  
Here the lower bound on $M_U$ follows
from proton decay.  Although for Coleman-Weinberg
potential the Planck constraint on $V_0$ corresponds to
$M_U\gtrsim4\times10^{16}$ GeV, lower $M_U$ values are possible for the smeared
Higgs potential. If the smearing parameter $x$ is close to zero, proton decay could 
also be observed in this class of models. 

Finally we note that $E_6$ breaking via $SU(3)_c \times SU(3)_L \times SU(3)_R$
can yield intermediate mass monopoles carrying three units of Dirac charge
\cite{Shafi:1984wk}.

\section*{Acknowledgements} Q.S. is supported in part by the DOE Grant DE-SC0013880 and thanks
Dylan Spence for reading the manuscript.

\bibliography{monopoles_v1}

\providecommand{\href}[2]{#2}\begingroup\raggedright\begin{thebibliography}{10}

\makeatletter
    \clubpenalty10000
    \@clubpenalty \clubpenalty
    \widowpenalty10000
\makeatother

\bibitem{Dirac:1931kp}
P.~A.~M. Dirac, ``{Quantized Singularities in the Electromagnetic Field},''
\href{http://dx.doi.org/10.1098/rspa.1931.0130}{{\em Proc. Roy. Soc. Lond.}
  {\bfseries A133} (1931) 60--72};\\
P.~A.~M. Dirac, ``{The Theory of magnetic poles},''
\href{http://dx.doi.org/10.1103/PhysRev.74.817}{{\em Phys. Rev.} {\bfseries 74}
  (1948) 817--830}.

\bibitem{Pati:1974yy}
J.~C. Pati and A.~Salam, ``{Lepton Number as the Fourth Color},''
  \href{http://dx.doi.org/10.1103/PhysRevD.10.275,
  10.1103/PhysRevD.11.703.2}{{\em Phys. Rev.} {\bfseries D10} (1974) 275--289}.
[Erratum: Phys. Rev.D11,703(1975)].


\bibitem{Georgi:1974sy}
H.~Georgi and S.~L. Glashow, ``{Unity of All Elementary Particle Forces},''
\href{http://dx.doi.org/10.1103/PhysRevLett.32.438}{{\em Phys. Rev. Lett.}
  {\bfseries 32} (1974) 438--441}.


\bibitem{'tHooft:1974qc}
G.~'t~Hooft, ``{Magnetic Monopoles in Unified Gauge Theories},''
\href{http://dx.doi.org/10.1016/0550-3213(74)90486-6}{{\em Nucl. Phys.}
  {\bfseries B79} (1974) 276--284};\\
A.~M. Polyakov, ``{Particle Spectrum in the Quantum Field Theory},'' {\em JETP
  Lett.} {\bfseries 20} (1974) 194--195.
[Pisma Zh. Eksp. Teor. Fiz.20,430(1974)].



\bibitem{Daniel:1979yz}
M.~Daniel, G.~Lazarides, and Q.~Shafi, ``{SU(5) Monopoles, Magnetic Symmetry
  and Confinement},''
\href{http://dx.doi.org/10.1016/0550-3213(80)90483-6}{{\em Nucl. Phys.}
  {\bfseries B170} (1980) 156}.

\bibitem{Lazarides:1980cc}
G.~Lazarides, M.~Magg, and Q.~Shafi, ``{Phase Transitions and Magnetic
  Monopoles in SO(10)},''
\href{http://dx.doi.org/10.1016/0370-2693(80)90553-5}{{\em Phys. Lett.}
  {\bfseries B97} (1980) 87}.

\bibitem{Lazarides:1984pq}
G.~Lazarides and Q.~Shafi, ``{Extended Structures at Intermediate Scales in an
  Inflationary Cosmology},''
\href{http://dx.doi.org/10.1016/0370-2693(84)91605-8}{{\em Phys. Lett.}
  {\bfseries B148} (1984) 35}.

\bibitem{Shafi:1983bd}
Q.~Shafi and A.~Vilenkin, ``{Inflation with SU(5)},''
\href{http://dx.doi.org/10.1103/PhysRevLett.52.691}{{\em Phys. Rev. Lett.}
  {\bfseries 52} (1984) 691--694}.

\bibitem{Shafi:1984tt}
Q.~Shafi and A.~Vilenkin, ``{Spontaneously Broken Global Symmetries and
  Cosmology},''
\href{http://dx.doi.org/10.1103/PhysRevD.29.1870}{{\em Phys. Rev.} {\bfseries
  D29} (1984) 1870}.

\bibitem{Parker:1970xv}
E.~N. Parker, ``{The Origin of Magnetic Fields},''
\href{http://dx.doi.org/10.1086/150442}{{\em Astrophys. J.} {\bfseries 160}
  (1970) 383}.

\bibitem{Hinshaw:2012aka}
{\bfseries WMAP} Collaboration, G.~Hinshaw {\em et~al.}, ``{Nine-Year Wilkinson
  Microwave Anisotropy Probe (WMAP) Observations: Cosmological Parameter
  Results},'' \href{http://dx.doi.org/10.1088/0067-0049/208/2/19}{{\em
  Astrophys. J. Suppl.} {\bfseries 208} (2013) 19},
\href{http://arxiv.org/abs/1212.5226}{{\ttfamily arXiv:1212.5226
  [astro-ph.CO]}}.

\bibitem{Ade:2015xua}
{\bfseries Planck} Collaboration, P.~A.~R. Ade {\em et~al.}, ``{Planck 2015
  results. XIII. Cosmological parameters},''
\href{http://arxiv.org/abs/1502.01589}{{\ttfamily arXiv:1502.01589
  [astro-ph.CO]}}.

\bibitem{Ade:2015lrj}
{\bfseries Planck} Collaboration, P.~A.~R. Ade {\em et~al.}, ``{Planck 2015
  results. XX. Constraints on inflation},''
\href{http://arxiv.org/abs/1502.02114}{{\ttfamily arXiv:1502.02114
  [astro-ph.CO]}}.

\bibitem{Shafi:2006cs}
Q.~Shafi and V.~N. Şenoğuz, ``{Coleman-Weinberg potential in good agreement
  with WMAP},'' \href{http://dx.doi.org/10.1103/PhysRevD.73.127301}{{\em Phys.
  Rev.} {\bfseries D73} (2006) 127301},
\href{http://arxiv.org/abs/astro-ph/0603830}{{\ttfamily arXiv:astro-ph/0603830
  [astro-ph]}}.

\bibitem{Rehman:2008qs}
M.~U. Rehman, Q.~Shafi, and J.~R. Wickman, ``{GUT Inflation and Proton Decay
  after WMAP5},'' \href{http://dx.doi.org/10.1103/PhysRevD.78.123516}{{\em
  Phys. Rev.} {\bfseries D78} (2008) 123516},
\href{http://arxiv.org/abs/0810.3625}{{\ttfamily arXiv:0810.3625 [hep-ph]}};\\
N.~Okada, V.~N. Şenoğuz, and Q.~Shafi, ``{The Observational Status of Simple
  Inflationary Models: an Update},''
\href{http://arxiv.org/abs/1403.6403}{{\ttfamily arXiv:1403.6403 [hep-ph]}}.

\bibitem{Linde:1981mu}
A.~D. Linde, ``{A New Inflationary Universe Scenario: A Possible Solution of
  the Horizon, Flatness, Homogeneity, Isotropy and Primordial Monopole
  Problems},''
\href{http://dx.doi.org/10.1016/0370-2693(82)91219-9}{{\em Phys. Lett.}
  {\bfseries B108} (1982) 389--393};\\
A.~Albrecht and P.~J. Steinhardt, ``{Cosmology for Grand Unified Theories with
  Radiatively Induced Symmetry Breaking},''
\href{http://dx.doi.org/10.1103/PhysRevLett.48.1220}{{\em Phys. Rev. Lett.}
  {\bfseries 48} (1982) 1220--1223}.

\bibitem{Guth:1980zm}
A.~H. Guth, ``{The Inflationary Universe: A Possible Solution to the Horizon
  and Flatness Problems},''
\href{http://dx.doi.org/10.1103/PhysRevD.23.347}{{\em Phys. Rev.} {\bfseries
  D23} (1981) 347--356}.

\bibitem{Coleman:1973jx}
S.~R. Coleman and E.~J. Weinberg, ``{Radiative Corrections as the Origin of
  Spontaneous Symmetry Breaking},''
\href{http://dx.doi.org/10.1103/PhysRevD.7.1888}{{\em Phys. Rev.} {\bfseries
  D7} (1973) 1888--1910}.

\bibitem{Albrecht:1984qt}
A.~Albrecht and R.~H. Brandenberger, ``{On the Realization of New Inflation},''
\href{http://dx.doi.org/10.1103/PhysRevD.31.1225}{{\em Phys. Rev.} {\bfseries
  D31} (1985) 1225}.

\bibitem{Linde:2005ht}
A.~D. Linde, ``{Particle physics and inflationary cosmology},'' {\em Contemp.
  Concepts Phys.} {\bfseries 5} (1990) 1--362,
\href{http://arxiv.org/abs/hep-th/0503203}{{\ttfamily arXiv:hep-th/0503203
  [hep-th]}}.

\bibitem{Smith:2008pf}
T.~L. Smith, M.~Kamionkowski, and A.~Cooray, ``{The inflationary
  gravitational-wave background and measurements of the scalar spectral
  index},'' \href{http://dx.doi.org/10.1103/PhysRevD.78.083525}{{\em Phys.
  Rev.} {\bfseries D78} (2008) 083525},
\href{http://arxiv.org/abs/0802.1530}{{\ttfamily arXiv:0802.1530 [astro-ph]}};\\
J.~Martin, C.~Ringeval, and V.~Vennin, ``{Encyclopædia Inflationaris},''
  \href{http://dx.doi.org/10.1016/j.dark.2014.01.003}{{\em Phys. Dark Univ.}
  {\bfseries 5-6} (2014) 75--235},
\href{http://arxiv.org/abs/1303.3787}{{\ttfamily arXiv:1303.3787
  [astro-ph.CO]}}.

\enlargethispage{1\baselineskip}

\bibitem{Lyth:2009zz}
D.~H. Lyth and A.~R. Liddle, {\em {The primordial density perturbation:
  Cosmology, inflation and the origin of structure}}.
\newblock 2009.
\newblock
\url{http://www.cambridge.org/uk/catalogue/catalogue.asp?isbn=9780521828499}.
\newblock

\bibitem{Liddle:2003as}
A.~R. Liddle and S.~M. Leach, ``{How long before the end of inflation were
  observable perturbations produced?},''
  \href{http://dx.doi.org/10.1103/PhysRevD.68.103503}{{\em Phys. Rev.}
  {\bfseries D68} (2003) 103503},
\href{http://arxiv.org/abs/astro-ph/0305263}{{\ttfamily arXiv:astro-ph/0305263
  [astro-ph]}}.

\bibitem{Rehman:2010es}
M.~U. Rehman and Q.~Shafi, ``{Higgs Inflation, Quantum Smearing and the Tensor
  to Scalar Ratio},'' \href{http://dx.doi.org/10.1103/PhysRevD.81.123525}{{\em
  Phys. Rev.} {\bfseries D81} (2010) 123525},
\href{http://arxiv.org/abs/1003.5915}{{\ttfamily arXiv:1003.5915
  [astro-ph.CO]}}.

\bibitem{Vilenkin:1994pv}
A.~Vilenkin, ``{Topological inflation},''
  \href{http://dx.doi.org/10.1103/PhysRevLett.72.3137}{{\em Phys. Rev. Lett.}
  {\bfseries 72} (1994) 3137--3140},
\href{http://arxiv.org/abs/hep-th/9402085}{{\ttfamily arXiv:hep-th/9402085
  [hep-th]}};\\
A.~D. Linde and D.~A. Linde, ``{Topological defects as seeds for eternal
  inflation},'' \href{http://dx.doi.org/10.1103/PhysRevD.50.2456}{{\em Phys.
  Rev.} {\bfseries D50} (1994) 2456--2468},
\href{http://arxiv.org/abs/hep-th/9402115}{{\ttfamily arXiv:hep-th/9402115
  [hep-th]}};\\
C.~Destri, H.~J. de~Vega, and N.~G. Sanchez, ``{MCMC analysis of WMAP3 and SDSS
  data points to broken symmetry inflaton potentials and provides a lower bound
  on the tensor to scalar ratio},''
  \href{http://dx.doi.org/10.1103/PhysRevD.77.043509}{{\em Phys. Rev.}
  {\bfseries D77} (2008) 043509},
\href{http://arxiv.org/abs/astro-ph/0703417}{{\ttfamily arXiv:astro-ph/0703417
  [astro-ph]}};\\
R.~Kallosh and A.~D. Linde, ``{Testing String Theory with CMB},''
  \href{http://dx.doi.org/10.1088/1475-7516/2007/04/017}{{\em JCAP} {\bfseries
  0704} (2007) 017},
\href{http://arxiv.org/abs/0704.0647}{{\ttfamily arXiv:0704.0647 [hep-th]}}.

\bibitem{Kibble:1976sj}
T.~W.~B. Kibble, ``{Topology of Cosmic Domains and Strings},''
\href{http://dx.doi.org/10.1088/0305-4470/9/8/029}{{\em J. Phys.} {\bfseries
  A9} (1976) 1387--1398}.

\bibitem{Ambrosio:2002qq}
{\bfseries MACRO} Collaboration, M.~Ambrosio {\em et~al.}, ``{Final results of
  magnetic monopole searches with the MACRO experiment},''
  \href{http://dx.doi.org/10.1140/epjc/s2002-01046-9}{{\em Eur. Phys. J.}
  {\bfseries C25} (2002) 511--522},
\href{http://arxiv.org/abs/hep-ex/0207020}{{\ttfamily arXiv:hep-ex/0207020
  [hep-ex]}}.

\bibitem{Kolb:1990vq}
E.~W. Kolb and M.~S. Turner, ``{The Early Universe},''
{\em Front. Phys.} {\bfseries 69} (1990) 1--547.

\bibitem{Adams:1993fj}
F.~C. Adams, M.~Fatuzzo, K.~Freese, G.~Tarle, R.~Watkins, and M.~S. Turner,
  ``{Extension of the Parker bound on the flux of magnetic monopoles},''
\href{http://dx.doi.org/10.1103/PhysRevLett.70.2511}{{\em Phys. Rev. Lett.}
  {\bfseries 70} (1993) 2511--2514}.

\bibitem{Ueno:2012md}
{\bfseries Super-Kamiokande} Collaboration, K.~Ueno {\em et~al.}, ``{Search for
  GUT monopoles at Super–Kamiokande},''
  \href{http://dx.doi.org/10.1016/j.astropartphys.2012.05.008}{{\em Astropart.
  Phys.} {\bfseries 36} (2012) 131--136},
\href{http://arxiv.org/abs/1203.0940}{{\ttfamily arXiv:1203.0940 [hep-ex]}};\\
{\bfseries IceCube} Collaboration, M.~G. Aartsen {\em et~al.}, ``{Search for
  non-relativistic Magnetic Monopoles with IceCube},''
  \href{http://dx.doi.org/10.1140/epjc/s10052-014-2938-8}{{\em Eur. Phys. J.}
  {\bfseries C74} no.~7, (2014) 2938},
\href{http://arxiv.org/abs/1402.3460}{{\ttfamily arXiv:1402.3460
  [astro-ph.CO]}}.

\bibitem{Callan:1982au}
C.~G. Callan, Jr., ``{Dyon-Fermion Dynamics},''
\href{http://dx.doi.org/10.1103/PhysRevD.26.2058}{{\em Phys. Rev.} {\bfseries
  D26} (1982) 2058--2068};\\
V.~A. Rubakov, ``{Adler-Bell-Jackiw Anomaly and Fermion Number Breaking in the
  Presence of a Magnetic Monopole},''
\href{http://dx.doi.org/10.1016/0550-3213(82)90034-7}{{\em Nucl. Phys.}
  {\bfseries B203} (1982) 311--348}.

\bibitem{Agashe:2014kda}
{\bfseries Particle Data Group} Collaboration, K.~A. Olive {\em et~al.},
  ``{Review of Particle Physics},''
\href{http://dx.doi.org/10.1088/1674-1137/38/9/090001}{{\em Chin. Phys.}
  {\bfseries C38} (2014) 090001}.

\bibitem{Dawson:1982sc}
S.~Dawson and A.~N. Schellekens, ``{Monopole Catalysis of Proton Decay in
  SO(10) Grand Unified Models},''
\href{http://dx.doi.org/10.1103/PhysRevD.27.2119}{{\em Phys. Rev.} {\bfseries
  D27} (1983) 2119};\\
E.~J. Weinberg, D.~London, and J.~L. Rosner, ``{Magnetic Monopoles With $Z(n$)
  Charges},''
\href{http://dx.doi.org/10.1016/0550-3213(84)90526-1}{{\em Nucl. Phys.}
  {\bfseries B236} (1984) 90};\\
A.~Sen, ``{Baryon Number Violation Induced by the Monopoles of the Pati-Salam
  Model},''
\href{http://dx.doi.org/10.1016/0370-2693(85)91441-8}{{\em Phys. Lett.}
  {\bfseries B153} (1985) 55}.

\bibitem{Nishino:2009aa}
{\bfseries Super-Kamiokande} Collaboration, H.~Nishino {\em et~al.}, ``{Search
  for Proton Decay via $p\to e^+ \pi^0$ and $p\to \mu^+ \pi^0$ in a Large Water
  Cherenkov Detector},''
  \href{http://dx.doi.org/10.1103/PhysRevLett.102.141801}{{\em Phys. Rev.
  Lett.} {\bfseries 102} (2009) 141801},
\href{http://arxiv.org/abs/0903.0676}{{\ttfamily arXiv:0903.0676 [hep-ex]}}.

\bibitem{Abe:2011ts}
K.~Abe {\em et~al.}, ``{Letter of Intent: The Hyper-Kamiokande Experiment ---
  Detector Design and Physics Potential ---},''
\href{http://arxiv.org/abs/1109.3262}{{\ttfamily arXiv:1109.3262 [hep-ex]}}.

\bibitem{Lee:1994vp}
D.-G. Lee, R.~N. Mohapatra, M.~K. Parida, and M.~Rani, ``{Predictions for
  proton lifetime in minimal nonsupersymmetric SO(10) models: An update},''
  \href{http://dx.doi.org/10.1103/PhysRevD.51.229}{{\em Phys. Rev.} {\bfseries
  D51} (1995) 229--235},
\href{http://arxiv.org/abs/hep-ph/9404238}{{\ttfamily arXiv:hep-ph/9404238
  [hep-ph]}}.

\bibitem{Acampora:1994rh}
F.~Acampora, G.~Amelino-Camelia, F.~Buccella, O.~Pisanti, L.~Rosa, and T.~Tuzi,
  ``{Proton decay and neutrino masses in SO(10)},''
  \href{http://dx.doi.org/10.1007/BF02787063}{{\em Nuovo Cim.} {\bfseries A108}
  (1995) 375--400},
\href{http://arxiv.org/abs/hep-ph/9405332}{{\ttfamily arXiv:hep-ph/9405332
  [hep-ph]}}.

\enlargethispage{1\baselineskip}

\bibitem{Shafi:1983gz}
Q.~Shafi and C.~Wetterich, ``{Modification of {GUT} Predictions in the Presence
  of Spontaneous Compactification},''
\href{http://dx.doi.org/10.1103/PhysRevLett.52.875}{{\em Phys. Rev. Lett.}
  {\bfseries 52} (1984) 875}.

\bibitem{Shafi:1984wk}
Q.~Shafi and C.~Wetterich, ``{Magnetic monopoles in grand unified and
Kaluza-Klein theories},'' in {\em Monopole '83, Ann Arbor, Proceedings}
47--49;\\
T.~W. Kephart and Q.~Shafi, ``{Family unification, exotic states and magnetic
  monopoles},'' \href{http://dx.doi.org/10.1016/S0370-2693(01)01187-X}{{\em
  Phys. Lett.} {\bfseries B520} (2001) 313--316},
\href{http://arxiv.org/abs/hep-ph/0105237}{{\ttfamily arXiv:hep-ph/0105237
  [hep-ph]}}.

\end{thebibliography}\endgroup

\end{document}